\documentclass[rnote]{aa} % for the research notes
\usepackage{natbib,graphics,txfonts}

\begin{document}

\title{A {\it Spitzer}-IRS view  of early-type galaxies with \\
cuspy/core nuclei and with fast/slow rotation} 

\author{R. Rampazzo\inst{1}
O. Vega\inst{2}, 
A. Bressan\inst{3}, 
M.S. Clemens\inst{1}, 
A. Marino\inst{1}
\and P. Panuzzo\inst{4}} 

\institute{INAF Osservatorio Astronomico di Padova, Vicolo
dell'Osservatorio 5, 35122 Padova, Italy\\
\email{roberto.rampazzo@oapd.inaf.it}
\and
Instituto Nacional de Astrof\'{\i}sica, Optica y Electr\'onica,
Apdos. Postales 51 y 216, C.P. 72000 Puebla, Pue., M\'exico
\and
Scuola Internazionale Superiore di Studi Avanzati (SISSA), via Bonomea, 
265 - 34136 Trieste ITALY
\and GEPI, Observatoire de Paris, CNRS, Univ. Paris Diderot,
Place Jules Janssen 92190 Meudon, France
}
  \authorrunning{Rampazzo et al.}
  \date{Received; accepted}
  
\abstract
{The recent literature suggests that an evolutionary dichotomy exists for early-type galaxies 
(Es and S0s, ETGs) whereby their central photometric structure ({\it cuspy} versus {\it core} 
central luminosity profiles),  and figure of rotation (fast (FR) vs. slow (SR) rotators), are determined 
by whether they formed by ``wet'' or ``dry'' mergers.} 
{We consider whether the mid infrared (MIR) properties of ETGs, with their sensitivity
to accretion processes in particular in the last few Gyr (on average $z\lesssim$0.2), can put further constraints on this picture.}
{We investigate a sample of 49 ETGs for which nuclear MIR properties and detailed 
photometrical and kinematical  classifications are available from the recent literature.}
{In the stellar light {\it cuspy/core} ETGs show a dichotomy that is mainly driven by 
their luminosity. However in the MIR, the brightest {\it core} ETGs show  evidence
that accretions  have triggered both AGN and star formation activity in the recent past, 
challenging a ``dry'' merger scenario.  
In contrast, we do find, in the Virgo and Fornax clusters, that {\it cuspy}  ETGs, 
fainter than M$_{K_s}=-24$, are predominantly passively evolving in the same epoch, 
while, in low density environments,  they tend to be more active.\\
A significant and statistically similar  fraction of  both FR (38$^{+18}_{-11}$\%) and 
SR (50$^{+34}_{-21}$\%) shows PAH features in their MIR spectra. 
Ionized and molecular gas are also frequently detected. 
Recent star formation episodes are then a common phenomenon in both
kinematical classes, even in those dominated by AGN activity, suggesting
a similar evolutionary path in the last few Gyr.}
{MIR spectra suggest that the photometric segregation  between 
{\it cuspy} and {\it core} nuclei and the dynamical segregation between 
FR and SR  must have originated before  $z\sim$0.2).}


\maketitle

\section{Introduction}

A relatively large fraction of ETGs at high-redshift show 
clear evidence of interaction and/or merger morphologies and active star formation
\citep[e.g.][]{Treu05} supporting the model view these galaxies
are produced by a halo merger process 
\citep[see e.g.][and reference therein]{Mihos04,Cox08,Khochfar11,DeLucia11}. 
A further element for high-redshift formation scenarios comes from their measured
[$\alpha$/Fe] ratios,  encoding  information about the time-scale of star formation. 
In massive ETGs this ratio has super-solar values, suggesting that they formed 
on relatively  short time-scales  \citep[see e.g][]{Chiosi98,Granato04,Thomas05,Annibali07,Clemens09}. 
\citet{Annibali07} estimated that a fair upper limit to  the recent 
{\it rejuvenation} episodes is  $\sim$25\% of the total
galaxy mass but that they are typically much less intense than that 
 \citep[see e.g. the {\it Spitzer}-IRS
study of NGC 4435 by][]{Panuzzo07}. 
However, rejuvenation signatures in ETGs are often detected, not only in the galaxy
nucleus, but also in the disk, rings and even in galaxy outskirts, as clearly shown by GALEX
\citep[e.g][]{Rampazzo07,Marino09,Salim10,Thilker10,Marino11},
so that the different phases of galaxy assembly/evolution, and their link
to morphological and kinematical signatures, are vivaciously debated.

The merger process may involve either relatively few  (major) or  multiple (minor) events
during the galaxy assembly. Furthermore, it  may or may not include dissipation
 (and star formation),  two possibilities often called 
``wet" and ``dry'' mergers, respectively \citep[see e.g.][]{vanDokkum05}.
Other mechanisms, however, like conversion of late-type galaxies into ETGs by environmental
effects, like strangulation, ram-pressure etc. \citep[e.g.][]{Boselli06}, and by energy 
feedback from supernovae may also be important \citep[e.g.][]{Kormendy09}. 

Two {\it observable}  quantities are thought to distinguish ETGs produced by
``wet'' and ``dry'' mergers.  The first, mainly fruit of 
high resolution observations with the {\it Hubble} Space Telescope and of high precision 
photometric analyses, is the presence of either a {\it cusp} or a {\it core} in the
 inner galaxy luminosity profile \citep{Lauer91,Lauer92,Cote06,Turner12}. In contrast to 
{\it cuspy} profiles, the surface brightness in {\it core} profiles becomes shallower as $r
\rightarrow 0$. The same concept is considered by \citet{Kormendy09} who divide ETGs
into {\it cuspy--core} and {\it core--less}, depending on whether the luminosity profile
{\it misses light} or has an {\it extra-light} component with respect to the extrapolation
of the Sersic's law at small radii. \citet{Kormendy09} suggest that {\it cuspy--core}
nuclei have been scoured by binary black holes (BHs) during (the last) dissipationless,
``dry'', major merger. In contrast, {\it core--less} nuclei originate from ``wet'' mergers.
Analogously,  \citet{Cote06} and \citet{Turner12} found an extra stellar nucleus in the 
profile decomposition of ETGs in their Virgo (ACSVCS) and Fornax (ACSFCS) surveys 
in addition to simple Sersic profiles, in particular in low-luminosity(/mass) ETGs. 
They proposed that the most important mechanism for the assembly
of a stellar nucleus is the infall of star clusters through dynamical friction, while for more 
luminous(/massive) galaxies a ``wet scenario'' (gas accretion by mergers/accretions 
and tidal torques) dominates.      

The second observable quantity is the kinematical class. The class is
defined by a parameter describing the specific baryonic angular momentum
defined as follows, $\lambda_r$=$\langle r|V|\rangle/\langle r \sqrt{V^2 +
\sigma^2}\rangle$, where $r$ is the galacto-centric distance, $V$ and
$\sigma$ are luminosity weighted averages of the rotation velocity and velocity
dispersion over a two-dimensional kinematical field. The measure refers to
the inner part of the galaxy, typically of the order or less than 1 effective radius, 
$r_e$, i.e. significantly larger than the regions where cusps and cores are 
detected. $\lambda_r$ divides ETGs into the two classes of fast (FR) and slow (SR) rotators
\citep[][and reference therein]{Emsellem11}. FR are by far the majority of ETGs
(86$\pm$2\% in the ATLAS$^{3D}$ survey).  SR represent massive ETGs that 
might have suffered from significant merging without being able to rebuild a fast rotating
 component. \citet[][]{Khochfar11}
 find that the underlying physical reason for the different growth histories is the 
 slowing down, and ultimately complete shut-down, of gas cooling in massive, SR galaxies.  
 On average, the last gas-rich major merger interaction in SR happens at $z > 1.5$, 
 followed by a series of minor mergers which build-up the outer layers of the remnant, 
 i.e. do not feed the inner part of the galaxy.

FRs in the models of \citet{Khochfar11} have different formation paths. The majority (78\%)
have bulge-to-total stellar mass ratios (B/T) larger than 0.5 and manage to grow stellar discs due
  to continued gas cooling as a results of frequent minor mergers. The remaining 22\% live in 
  high--density environments and consist of low B/T galaxies with gas fractions 
  below 15\%, that have exhausted their cold gas reservoir and have no hot halo 
  from which gas can cool. 
Summarizing, a dissipative merging and/or a gas accretion episode from interacting
companions,  could be the way for the galaxy to rebuild a fast-rotating disk-like
component.  SR and FR basically correspond to the paradigms of ``dry'' vs. ``wet'' 
accretions/mergers respectively. 

Recently, \citet{Lauer12} attempted to unite the structural and kinematical views,
claiming that they are the two aspects of the same process.  Using the specific
angular momentum $\lambda_{r_e/2}$, computed from the 2D kinematics
within  half the effective radius by \citet{Emsellem11}, \citet{Lauer12} showed that 
{\it core} galaxies have rotation amplitudes  $\lambda_{r_e/2} \leq 0.25$ while all galaxies
with $\lambda_{r_e/2} > 0.25$ and ellipticity $\epsilon_{r_e/2} > 0.2$ lack cores.
Some FR have a core profile but they argue that both figure of rotation and the 
central structure of ETGs should be used together to separate systems that appear 
to have formed from ``wet'' and ``dry'' mergers. \citet{Krajnovic13b} show, however, 
that there is a genuine population of FR with cores. They suggest that the cores of both FR 
and SR are made of old stars and are found in galaxies typically lacking  molecular
and atomic gas, with few exceptions.

For the sake of simplicity throughout the paper, we will call {\it core}
 ETGs those galaxies for which the luminosity profile shallows out as 
 $r \rightarrow 0$ (i.e. cuspy-core in \citet{Kormendy09},
 core in \citet{Lauer12}, non-nucleated in \citet{Cote06,Turner12}). 
 We will refer to {\it cuspy} ETGs as those which present an
 extra central luminosity component (i.e. core-less in \citet{Kormendy09},
 power-law + intermediate in \citet{Lauer12}, nucleated in 
\citet{Cote06,Turner12}) with respect to a fit of a Sersic model.
Depending on the accurate surface brightness profile decomposition performed by the 
above authors, rarely  the {\it cuspy} versus {\it core} classification given by
different authors for the same ETG is discrepant.

This note aims to contribute to the debate on the origin of {\it core}/{\it cuspy} 
and FR/SR ETGs, and the connection to the ``wet'' vs. ``dry'' merger hypotheses,  
using mid-infrared (MIR) spectra of well-studied ETGs. 
The paper is organized as follows. In \S~2 we briefly describe how {\it Spitzer}-IRS
spectra trace the recent few Gyr evolution in ETGs.   
We present the MIR  vs. the  {\it cuspy}/{\it core} nuclear properties of ETGs 
\citep{Kormendy09,Cote06,Turner12,Lauer12} in \S~3, 
and vs. the FR/SR kinematical classes  \citep{Emsellem11}
in \S~4. Conclusions are presented in \S~5.

%------------------------- begin Table 1 -----------------------------------
\begin{table*}
 \begin{minipage}{180mm}
  \caption{The Virgo sample in \citet{Kormendy09} with {\it Spitzer}-IRS MIR class}
  \begin{tabular}{llccccccc}
  \hline
 Galaxy &  RSA   & n$_{tot}$ &  \%               & $M_{K_s}$ & MIR   & Kinematical & Kinematical and Morphological & Dust\\
             &  Type    &     &  light&  &  class & class       & peculiarities  &\\
             &               &       &                      &               &                              &                                      &   & \\
 \hline            
core &               &               &  missing light &   &   & & \\
\hline
NGC 4472 & E1/S0$_1$(1) & 5.99$^{+0.31}_{-0.29}$  & -0.50$\pm$0.05 & -25.73 & 1 &SR & CR s-s (1) & Y\\
NGC  4486 & E0 & 11.84$^{+1.79}_{-1.19}$ & -4.20$\pm$1.00  & -25.31 &  4&  SR & SC(2), jet & Y\\
NGC 4649 & S0$_1$(2) & 5.36$^{+0.38}_{-0.32}$ & -1.05$\pm$0.07 & -25.35 & 1 & FR  & asym. rot. curve (6) & N\\
NGC 4365 & E3 & 7.11$^{+0.40}_{-0.35}$  & -0.63$\pm$0.07 & -25.19 & 0 & SR & Faint SW fan (4) & N \\
NGC 4374 & E1 & 7.98$^{+0.71}_{-0.56}$ & -1.52$\pm$0.05  & -25.13 & 2 & SR &V$\approx$0 (3); SC (2) & Y \\ 
NGC 4261 & E3 &  7.49$^{+0.82}_{-0.60}$ & -1.84$\pm$0.05  & -25.24 & 4& SR  &  NW tidal arm, faint SE fan (4) & Y\\
NGC 4382 & S0$_1$(3) pec & 6.12$^{+0.31}_{-0.27}$ & -0.18$\pm$0.06 & -25.13 & 1 &FR& MC(2), shells  (7) & Y \\ 
NGC 4636 & E0/S0$_1$(6) & 5.65$^{+0.48}_{-0.28}$ & -0.22$\pm$0.04 & -24.42 & 2& SR & gas irr. motion (a) & Y\\
NGC 4552 & S0$_1$(0)  & 9.22$^{+1.13}_{-0.83}$ & -1.23$\pm$0.09 & -24.31 & 2 & SR &  KDC (2)  shells (4) & Y \\
&             &            &   & &   &      &  \\
\hline
cuspy &           &          &  extra light   &     & &    & \\
\hline
&             &            &   &  & &      &  \\
NGC 4621 & E5 &5.36$^{+0.30}_{-0.28}$  & 0.27$\pm$0.06 & -24.13 & 0 & FR&  KDC (2)  & N\\
NGC 4473 & E5 & 4.00$^{+0.18}_{-0.16}$ & 8.80$\pm$1.00 & -23.77  &  0 & FR & MC (2) &N \\
NGC 4478 & E2  & 2.07$^{+0.08}_{-0.07}$ & 1.12$\pm$0.15 & -23.15 & 0&FR &  \dots & N\\
NGC 4570 & S0$_1$(7)/E7 &  3.69$\pm$0.50 & \dots  & -23.49 & 0 & FR & MC (2) & N\\
NGC 4660 & E5 & 4.43$\pm$0.38 & \dots & -22.69 & 0 & FR &  MC (2) & N\\
NGC 4564 & E6 & 4.69$\pm$0.20 & \dots & -23.09 & 0  & FR & SC (2)  &  N\\
 \hline
\hline
\end{tabular}
\label{tab1}

Notes: {\it Core} and {\it cuspy} galaxies correspond to cuspy-core and core-less, respectively,
in \citet{Kormendy09}. 
The Sersic's index, $n_{tot}$, and the percentage of extra light  are taken from \citet{Kormendy09}.
The $M_{K_s}$ absolute magnitude and the MIR class is obtained from \citet{Brown11} and \citet{Rampazzo13}. 
Kinematical classes are from \citet{Emsellem11}. Kinematical and morphological peculiarities 
are coded as follows: {CR s-s}: counter rotation stars vs. stars (1) 
\citet{Corsini98}; KDC indicates a kinematically decoupled 
component, not necessarily counter-rotation; MC multiple 
components; SC single component \citet[][]{Krajnovic08} 
(2). A description of the kinematic and morphological peculiarities  
of the galaxies and full references are reported in: 
(3) \citet[][]{Emsellem04}; (4) \citet{Tal09}; (5) \citet{MC83};  
(6) \citet{Pinkney03}; (7) \citet{Kormendy09}.
Dust properties (Y if present) are taken from \citet{Cote06} and Table B3 in \citet{Rampazzo13}.
\end{minipage}
\end{table*}
%--------------------------------- end Table 1 -------------------------------

\section{A classification of {\it Spitzer}-IRS spectra of ETGs}

Using {\it Spitzer}-IRS, \citet{Rampazzo13} have recently produced an 
atlas of low resolution MIR spectra, of  91 nearby 
(D$\leq$72 Mpc) ETGs in the {\it Revised Shapley-Ames Catalog} 
(RSA). Spectra are extracted within a common aperture (3.6\arcsec$\times$18\arcsec)
providing an integral view of the galaxy's inner region. On average, the area 
covered by the rectangular IRS  aperture is about 2-3 
times the area of a circle of radius $r_e/8$. MIR spectra then cover a region where the
{\it cuspy}/{\it core} profiles and/or the departures from the Sersic's profile are measured, 
as well as a significant part of the region where  FR vs. SR are separated, using e.g.
$\lambda_{r_e/2}$. In addition, ``dry''  vs. ``wet'' mergers {\it maximize} their possible
signatures in the galaxy nucleus, as shown by the study of line-strength indices 
\citep[e.g.][and references therein]{Annibali07}, by counter-rotating components 
\citep[e.g.][]{Emsellem11} and morphological peculiarities such as irregular/chaotic 
dust-lanes \citep[e.g.][]{Cote06,Turner12}.
 
For each galaxy, the atlas provides the fully reduced and calibrated spectrum, 
the intensity of nebular and molecular emission lines and Polycyclic 
Aromatic Hydrocarbons (PAHs), after a template spectrum of a passively evolving 
ETG has been subtracted. Spectra are  classified into five mid-infrared classes,
ranging from AGN (class-4) and star forming nuclei
(class-3), transition class-2 (with PAHs) and class-1 (no-PAHs)
 to passively evolving nuclei (class-0) \citep{Bressan06}.  
\citet{Panuzzo11} suggest that each of the five MIR 
classes is a snapshot of the evolution of ETG nuclei during an accretion episode, 
starting from, and ending with, a class-0 spectrum (see their Figure~11). 
\citet{Vega10} suggest that anomalous PAHs in class-2 spectra are produced 
by carbon stars which are present in stellar populations with 
ages in the range of a few Gyr  \citep[][see their Table~1]{Nanni13}, depending
on the metallicity, corresponding, on average, to a redshift coverage $z\lesssim 0.2$. 

%---------------------------- figure 1 ---------------------------------------------
\begin{figure*}
\centering
\includegraphics[width=12.5cm]{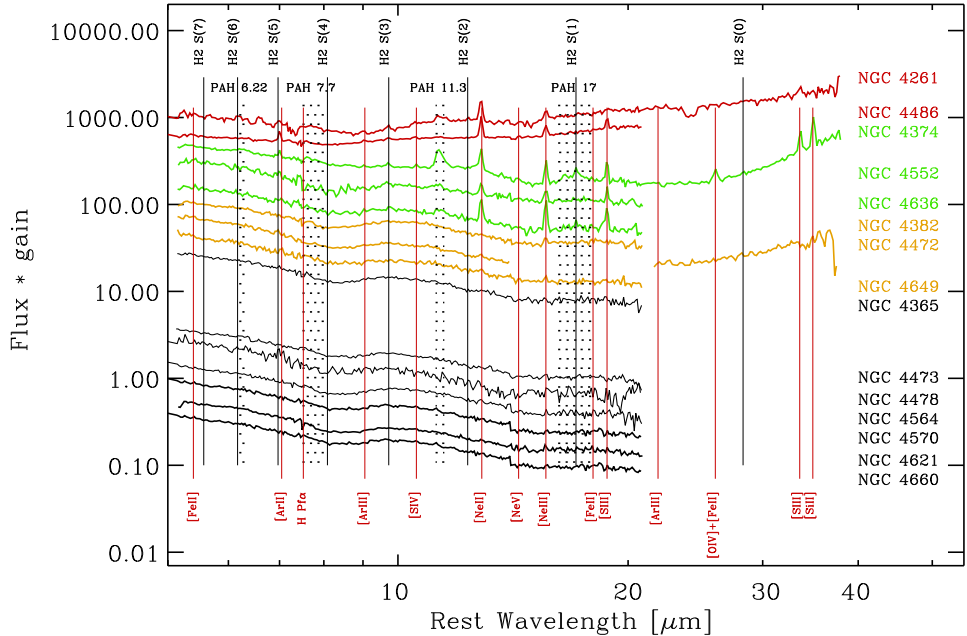}
\caption{ ETGs in the sample of \citet{Kormendy09}. MIR classes are
colour coded (black = class~0; yellow = class~1; green = class~2 and red = class~4.
Nebular lines in the spectra are indicated with vertical red solid lines.
PAH complexes are indicated with dotted areas. H$_2$ rotational emission lines are
indicated with black solid lines.}
\label{fig1}
\end{figure*}
%-------------------------- end figure1  ----------------------------------------------------

MIR spectral classes show a dichotomy between the Es and S0s. i.e.
the classical morphological classification provided by RSA. 
\citet{Rampazzo13} found that 46$^{+11}_{-10}$\% of Es 
and 20$^{+11}_{-7}$\% of S0s have a  spectrum of class-0, 
i.e.  Es are significantly  more passive than S0s. 
A small, similar  ($\sim$9$^{+4}_{-3}$\%) fraction of Es and S0s  have
PAH ratios  typical  of  star forming galaxies. PAHs are detected in
 47$^{+8}_{-7}$\% of ETGs and with a statistically similar fraction in Es and S0s.
Basically half of ETGs (41$^{+8}_{-7}$\%) in the sample show kinematical and/or 
 morphological scars of recent accretion episodes. 
 %Even class-0, i.e. passively evolving 
 %ETGs, show such signatures, supporting  the view of an evolutionary link 
 %between the MIR classes and accretion/feedback  phenomena.

In this note we investigate ETGs subdivided into the new, well defined photometric 
({\it cuspy vs. core}) and kinematical (FR/SR) classes. 
To this end, we cross-correlate the MIR RSA sample 
described above with the \citet{Kormendy09},  \citet{Lauer12}, \citet{Cote06}, 
\citet{Turner12} and ATLAS$^{3D}$ samples 
\citep[see e.g.][and references therein]{Emsellem11} to investigate the recent 
nuclear star formation history.  14 ETGs in Virgo, in particular 
the brighter ones,  are in common with \citet{Kormendy09} who classified 
them as either cuspy-core or core-less (i.e. {\it core} and {\it cuspy} 
in the present paper). 26 are in common with the
ACSVCS+ACSFCS  surveys in which galaxies are separated into
core+Sersic and Sersic type nuclei \citep{Cote06,Turner12}. 
41 out of 91 ETGs in the RSA MIR atlas are in common with the 
ATLAS$^{3D}$ survey sample.  For these galaxies \citet{Emsellem11} provide 
a classification into FR or SR. The inner luminosity profiles of these ETGs have been
 classified into {\it cuspy} (power-law and intermediate) or 
 {\it core} by \citet{Lauer12} (23 out of  41) and by \citet{Krajnovic13b}
 for a total of 44 objects.

\section{MIR spectra vs. nuclear features in ETGs}

\subsection{The Virgo sample of \citet{Kormendy09}}

The structural and photometric properties of a sample of Es (some
galaxies are classified S0  in the Revised Shapley Ames Catalogue
\citep{Sandage87}  as shown in Table~\ref{tab1}) in the Virgo cluster  have been well
studied by \citet{Kormendy09}. The authors divided their  sample
according to departures, at small radii, from the Sersic, $log I \propto
r^{1/n}$, law. 10 Es with total absolute magnitudes $M_{VT}\leq-21.66$, 
i.e. the brightest objects of their sample, have {\it core} luminosity profiles. 
This class of ETGs {\it miss light} with respect to the  extrapolation of the 
Sersic's law at small radii.
17 fainter galaxies, mostly Es in RSA, in the magnitude range $-21.54 \leq
M_{VT} \leq-15.53$, are {\it cuspy}, having excess light in their centre
with respect to the inward extrapolation of the outer Sersic profile. 
 
With the exception of NGC~4406, all {\it core} ETGs in the \citet{Kormendy09} sample, 
listed in the top part of Table~\ref{tab1}, have a MIR spectrum in the RSA sample 
of \citet{Rampazzo13}. IRS spectra are also available for three {\it cuspy} ETGs, namely
NGC 4621, NGC 4473 and NGC 4478.
NGC~4570, NGC 4660 and NGC 4564, having  absolute magnitudes 
similar to that of NGC~4478, also have IRS spectra but values for the extra light
component have not been quantified by \citet{Kormendy09}. 
We collect the basic data  in Table~\ref{tab1}.
 
According to \citet{Kormendy09}, {\it core} and {\it cuspy} ETGs have 
fundamentally different properties. {\it Core} ETGs, are slowly rotating, have
anisotropic velocity distributions, boxy isophotes, Sersic index $n>4$
and are $\alpha$-element enhanced, i.e. their stellar populations formed
on short timescales. In contrast, {\it cuspy} ETGs rotate rapidly, are
more isotropic and have disky isophotes.
They tend to have lower values of the Sersic index, $n\simeq 3 \pm 1$, and
 be less $\alpha$-enhanced. 
All {\it cuspy} ETGs are fast rotators, while the majority (7 out of 9) of {\it core} Es 
are slow rotators within $r_e/2$ \citep{Emsellem11} (Table~\ref{tab1}, column 7).

%----------------------  Table 2 -------------------------------------------
\begin{table*}
 \begin{minipage}{180mm}
 \begin{center}
\caption{The sample of ATLAS$^{3D}$ ETGs with a {\it Spitzer}-IRS class}
\begin{tabular}{llcccccccc}
\hline
 Ident. & Morpho.     & T   & P  & $\lambda_{r_e/2}$ & $\epsilon_{r_e/2}$ & V/$\sigma_{r_e/2}$ &  FR/SR$_{r_e/2}$ & $n_{tot}$ & MIR\\
              & RSA              &      &       &                                         &                                     &                                         & class                         &                    &class\\
\hline
NGC  821 & E6 & -4.8$\pm$0.4  & cuspy & 0.27  & 0.39 & 0.29 & FR & 10.4$\pm$0.7 & 0\\
NGC 2685 & S0$_3$(7) pec & -1.0$\pm$0.8 & cuspy & 0.63 & 0.19 & 0.73 & FR &  4.3$\pm$0.4 & 2 \\
NGC 2974 & E4 & -4.2$\pm$1.2 & cuspy & 0.66 & 0.40 &  0.81 & FR & 4.0$\pm$0.4  & 2 \\
NGC 3193 & E2 & -4.8$\pm$0.5  & core   & 0.20  & 0.14 & 0.21 & FR & 5.3$\pm$0.1 & 0 \\
NGC 3245 & S0$_1$ & -2.1$\pm$0.5  & cuspy & 0.59 & 0.44 & 0.58 & FR & 3.2$\pm$1.9 & 3 \\
NGC 3377 & E6 & -4.8$\pm$0.4 & cuspy & 0.52 & 0.50 &0.56 &  FR & 5.0$\pm$0.5 & 0 \\
NGC 3379 & E0 &-4.8$\pm$0.5  & core & 0.16 & 0.10 &  0.15 & FR & 5.3$\pm$0.9 & 0 \\ 
NGC 3608 &  E1 & -4.8$\pm$0.5 &core & 0.04 &0.19 & 0.05 & SR  & 3.9$\pm$0.5 & 0  \\ 
NGC 4036 & S0$_3$(8)/Sa &-2.6$\pm$0.7 &$\dots$ & 0.68 & 0.54 & 0.80 & FR & 2.0$\pm$0.1 & 2\\
NGC 4261 & E3 & -4.8$\pm$0.4  & core & 0.09 &0.22 &  0.09 & SR  &  5.1$\pm$0.4 & 4  \\
NGC 4339 & S0$_{1/2}(0)$ &-4.7$\pm$0.8 & cuspy & 0.31 & 0.06 & 0.30 & FR & 4.1$\pm$0.6 & 0 \\
NGC 4365 & E3 & -4.8$\pm$0.5  & core & 0.09 & 0.25 & 0.11 & SR & 5.2$\pm$0.4 & 0 \\
NGC 4371 & SB0$_{2/3}$(r)(3) & -1.3$\pm$0.6& cuspy  & 0.48 &0.31 & 0.54 & FR & 3.8$\pm$0.6 & 2 \\
NGC 4374 & E1 & -4.3$\pm$1.2  & core & 0.02 &0.15  & 0.03 & SR & 6.0$\pm$0.3 &  2 \\ 
NGC 4377 & S0$_1$(3) &-2.6$\pm$0.6  & cuspy & 0.34 & 0.23 & 0.31 & FR & 2.2$\pm$1.2 & 0\\
NGC 4382 & S0$_1$(3) pec &-1.3$\pm$0.6 & core &0.16 & 0.20 & 0.17 & FR & 5.1$\pm$1.3 & 1 \\
NGC 4435 & SB0$_1$(7) & -2.1$\pm$0.5  & $\dots$ & 0.60 & 0.46 & 0.67 & FR & 4.7$\pm$0.3 & 3 \\
NGC 4442 & SB0$_1$(6) &-1.9$\pm$0.4  & cuspy & 0.34 & 0.31 & 0.33 & FR & 2.8$\pm$0.2 & 0\\
NGC 4472 & E1/S0$_1$(1)&-4.8$\pm$0.5   & core & 0.08 & 0.17 & 0.07 & SR &  4.7$\pm$0.1 & 1\\
NGC 4473 & E5 & -4.7$\pm$0.7  & core & 0.25  & 0.40 &  0.26 & FR & 5.7$\pm$0.5 &0\\
NGC 4474 & S0$_1$(8) &-2.0$\pm$0.5     & cuspy &0.35 & 0.47 &  0.35 & FR & 3.5$\pm$0.4 & 2\\
NGC 4477 & SB0$_{1/2}$/SBa &-1.9$\pm$0.4 & cuspy & 0.22 & 0.13 & 0.21 & FR  & 4.1$\pm$1.2 & 2\\ 
NGC 4478 & E2 & -4.8$\pm$0.4  & core & 0.18 & 0.17 & 0.17 & FR & 2.0$\pm$0.1 & 0 \\
NGC 4486 & E0 &-4.3$\pm$0.6  & core  & 0.02  & 0.04 &  0.02 & SR & 2.9$\pm$0.2 &4 \\ 
NGC 4550 & E7/S0$_1$(7) &-2.1$\pm$0.7   & cuspy & 0.06 & 0.63 & 0.07 & SR & 1.7$\pm$0.1 & 3 \\
NGC 4552 & S0$_1$(0)  & -4.6$\pm$0.9 & core & 0.05 & 0.05 & 0.05 & SR & 6.2$\pm$0.4 & 2 \\
NGC 4564 & E6 & -4.8$\pm$0.5  & cuspy & 0.54  & 0.48 &  0.53 & FR & 2.9$\pm$0.2 &0 \\
NGC 4570 & S0$_1$(7)/E7 & -2.0$\pm$0.7  & cuspy & 0.50 & 0.55 & 0.47 & FR & 2.4$\pm$0.2 &0 \\
 NGC 4621 & E5 & -4.8$\pm$0.5  & cuspy & 0.29 & 0.36 & 0.27 & FR & 4.3$\pm$0.2 & 0 \\
NGC 4636 & E0/S0$_1$(6) & -4.8$\pm$0.5  & core & 0.04 & 0.09 & 0.04 & SR & 5.5$\pm$0.5 & 2\\
NGC 4649 & S0$_1$(2) &-4.6$\pm$0.8 & core & 0.13 & 0.16 & 0.12 & FR & 5.1$\pm$0.5 &1 \\
NGC 4660 & E5 & -4.7$\pm$0.5  & cuspy & 0.47 & 0.32 &  0.52 & FR  & 3.5$\pm$0.2 & 0\\
NGC 4697 & E6 & -4.4$\pm$0.8  & cuspy & 0.47 & 0.32 &   0.36 & FR & 4.6$\pm$0.2 &  3\\
NGC 5273 & S0/a & -1.9$\pm$0.4 & cuspy & 0.48 & 0.12 & 0.51 & FR & 1.8$\pm$1.1 & 4 \\
NGC 5353 & S0$_1$(7)/E7 & -2.1$\pm$0.6  & $\dots$ & 0.53 & 0.54 & 0.54 & FR & 3.3$\pm$0.5 & 2 \\
NGC 5631 & S0$_3$(2)/Sa &-1.9$\pm$0.4 & $\dots$ & 0.17 & 0.17  & 0.19 & FR & 4.3$\pm$0.5 & 2 \\
NGC 5638 & E1 & -4.8$\pm$0.4  & $\dots$ & 0.23 & 0.08 & 0.22 & FR & 3.5$\pm$0.2 & 0 \\
NGC 5813 & E1 & -4.8$\pm$0.4  & core & 0.07 & 0.17 &  0.16 & SR & 5.8$\pm$1.7 &1\\ 
NGC 5831 & E4 & -4.8$\pm$0.5 & cuspy & 0.06 & 0.20 & 0.16 & SR & 4.3$\pm$0.2  & 0 \\
NGC 5846 & S0$_1$(0) &-4.7$\pm$0.7  & core & 0.03 & 0.06 & 0.04 & SR & 3.9$\pm$0.2 & 1 \\
NGC 7332 & S0$_{2/3}$ & -1.9$\pm$0.5& cuspy & 0.34 & 0.44 & 0.29 & FR & 2.3$\pm$0.4 & 1\\
\hline
NGC 1339 & E4 & -4.3$\pm$0.5& cuspy & $\dots$ & $\dots$ & $\dots$ & $\dots$ & $\dots$ & 0\\
NGC 1351 & S0$_1$(6)/E6 &-3.1$\pm$0.6   & core & $\dots$ & $\dots$ & $\dots$ & $\dots$ & $\dots$ & 0\\
NGC 1374 & E0 &-4.4$\pm$1.1  & core & $\dots$ & $\dots$ & $\dots$ & $\dots$ & $\dots$ & 0\\
 NGC 1389 & S0$_1$(5)/SB0$_1$ &-2.8$\pm$0.7 & cuspy & $\dots$ & $\dots$ & $\dots$ & $\dots$ & $\dots$ & 0\\
NGC 1399 &  E1 & -4.6$\pm$0.5 & core & $\dots$ & $\dots$ & $\dots$ & $\dots$ & $\dots$ & 0\\
NGC 1404 & E2 & -4.8$\pm$0.5  & core & $\dots$ & $\dots$ & $\dots$ & $\dots$ & $\dots$ & 1\\
 NGC 1427 & E5 & -4.0$\pm$0.9  & core & $\dots$ & $\dots$ & $\dots$ & $\dots$ & $\dots$ & 0\\
 IC 2006  & E1 & -4.2$\pm$0.9  & cuspy & $\dots$ & $\dots$ & $\dots$ & $\dots$ & $\dots$ & 1\\
 \hline
\end{tabular}
  \end{center}
  ETGs in ATLAS$^{3D}$ (top part of the table), including the Virgo sample, plus Fornax cluster ETGs 
  \citep{Turner12} represent the global sample used for the {\it core} vs. {\it cuspy} and FR vs. SR analyses.
Morphological classifications in columns 2 and 3 are from RSA and {\tt LEDA}, respectively. For the
ATLAS$^{3D}$ sample, the profile type (column 4), P, is {\it cuspy} (originally $\setminus$=power law, $\wedge$ = intermediate) and {\it core} (originally $\cap$) is taken from  \citet{Krajnovic13b}.
The specific angular momentum (column 5), $\lambda_{r_e/2}$, the  ellipticity (column 6), $\epsilon_{e/2}$,  the rotation velocity to velocity dispersion ratio (column 7), V/$\sigma_{r_e/2}$ and the kinematical class (column 8),
 FR/SR,  at r$_e$/2 are from \citet{Emsellem11}. The Sersic index (column 9), $n_{tot}$, 
is from \citet{Krajnovic13a}. The MIR class (column 10) is taken from \citet{Rampazzo13}.  
\label{tab2}
\end{minipage}
\end{table*}
%-------------------------------------- end Table 2 ---------------------------------------------

The MIR spectra of the sub-set of \citet{Kormendy09} ETGs is shown in Figure~\ref{fig1}.   
All {\it cuspy} galaxies have a passive, class-0 spectrum characterized by the 
10~$\mu$m silicon feature produced by oxygen rich AGB stars. 
Their kinematics show signatures of past interaction/accretion episodes, 
such as kinematically decoupled cores (KDC) and multiple velocity components 
(MC). All have low nuclear X-ray luminosities 
(38.18 $\leq$ log $L_{X,nucl} \leq$ 38.92 erg s$^{-1}$, \citet{Pellegrini10})
indicating that no AGN are active, as also supported by the low radio power at 1.4 GHz
(most are upper limits while NGC 4563 has P$_{1.4 GHz}$=4.9 10$^{19}$ W Hz$^{-1}$, \citet{Brown11}) \citep[see also][Table B1]{Rampazzo13}.

A more active picture emerges for {\it core} ETGs.  The MIR spectra indicate
nuclei with different kinds of activity  from AGN (class-4, M87 and NGC 4261) to 
post-star forming  (MIR class-2 with anomalous PAHs, NGC 4374, 
NGC 4636, NGC 4552) to the quiescent nucleus of NGC 4365 (class-0). 

Both SR, the majority, and FR show signatures of kinematical and morphological 
peculiarities suggesting the action of accretion episodes.
Of particular interest is NGC~4382, which although showing strong shell
features (see e.g. figure 9 in \citet{Kormendy09}) has a nearly
passively evolving nucleus (class-1). This is not
the case for NGC~4374 and NGC~4552, that both show fine structure 
\citep[see][]{Sansom00}, although less prominent than that in NGC~4382.
The class-2 spectrum of NGC~4552 reveals the presence of gas and the
additional presence of anomalous PAHs, as in NGC 4374, points to the
occurrence of a recent star formation event. 
 
Summarizing, there is evidence that accretions in brightest, {\it core} Es in the 
\citet{Kormendy09} sample have triggered both AGN and SF activity, 
a challenge for completely dry mergers/accretions in the last  few Gyrs.
In contrast, the few {\it cuspy} galaxies in the present MIR sample are all passively evolving.
They may have either be quiescent for a very long time (the radio power may take longer to 
vary \citep[see e.g.][]{Brown11})  or possible accretions have been ``sterile'', activating
 neither AGN nor star formation.
 %at odd from what is expected from a ``wet'' accretion.
 An additional hint of a different recent evolution of the two photometric classes 
 may come from the observation that dust appears to be totally absent in 
 {\it cuspy} ETGs, while most of the {\it core} ETGs do show dust (col. 9 in Table 1). 
 \citet{Clemens10} found that dust can survive no longer than a few tens of Myr in
  the hostile ETG environment, so if present, should be quite recently acquired. 

%---------------------------- figure 2 ---------------------------------------------
\begin{figure*}
\centering
\includegraphics[width=8.8cm]{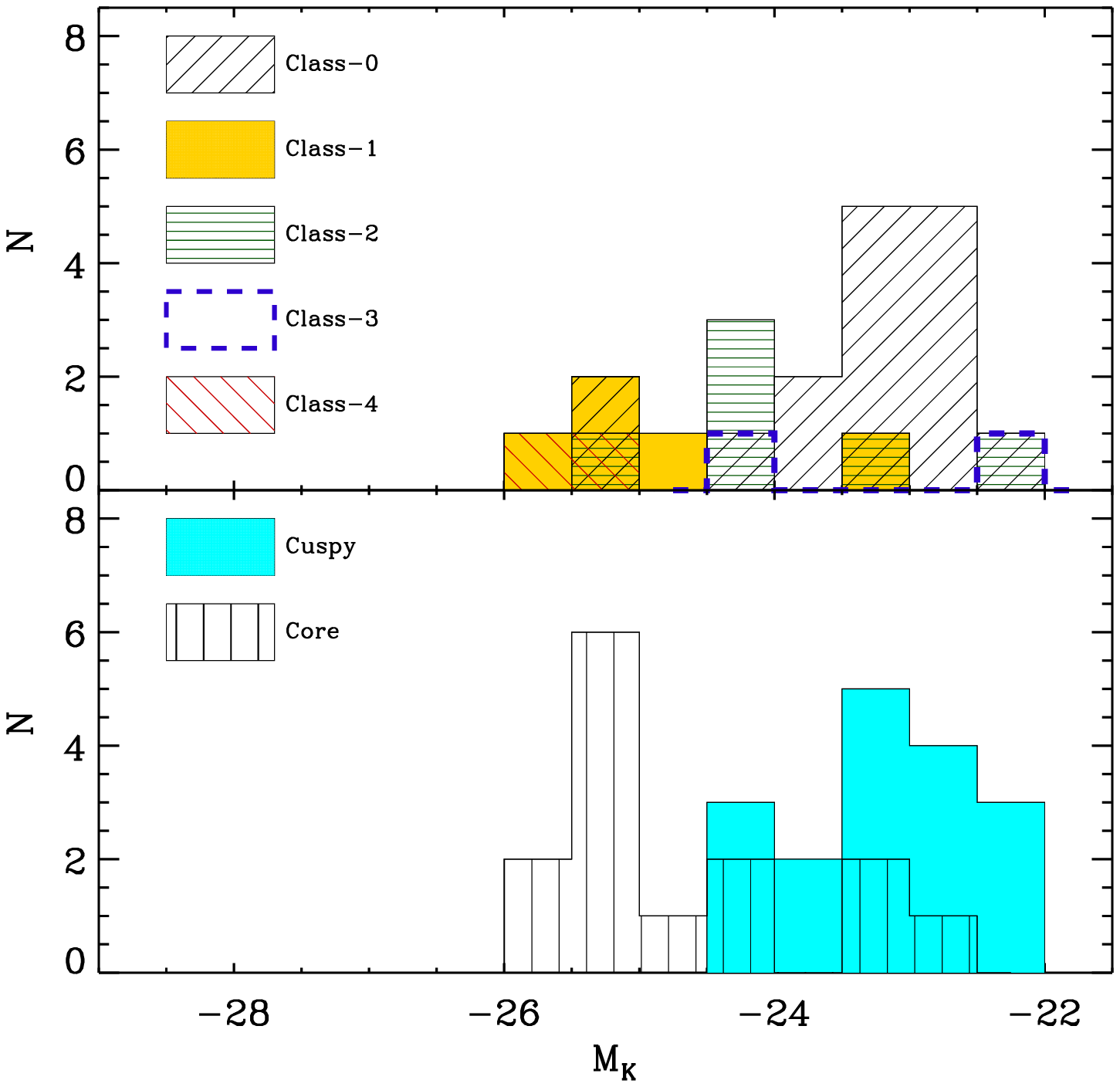}
\includegraphics[width=8.8cm]{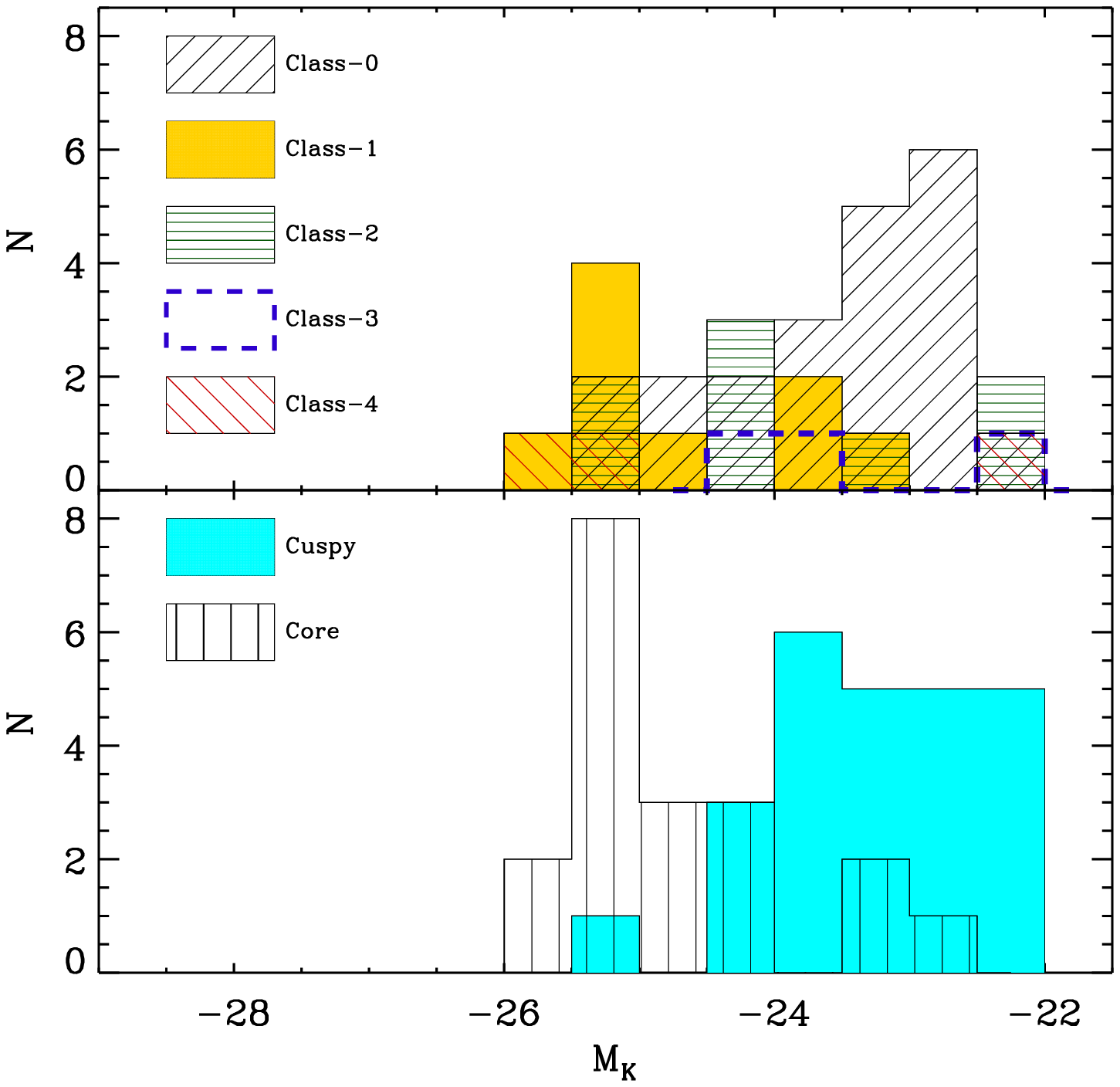}
\medskip \medskip \medskip \medskip
\caption{ K$_s$-band absolute magnitude, $M_K$, distribution vs. MIR classes (top panels)
and {\it cuspy}/{\it core} ETGs in the Virgo plus Fornax clusters \citep{Cote06,Turner12} (left panel)
and in the total sample, including ETGs in low density environments in \citet{Krajnovic13b}.} 
\label{fig2}
\end{figure*}
%-------------------------- end figure2  ----------------------------------------------------

%---------------------------- figure 3 ---------------------------------------------
\begin{figure*}
\centering
\includegraphics[width=12.5cm]{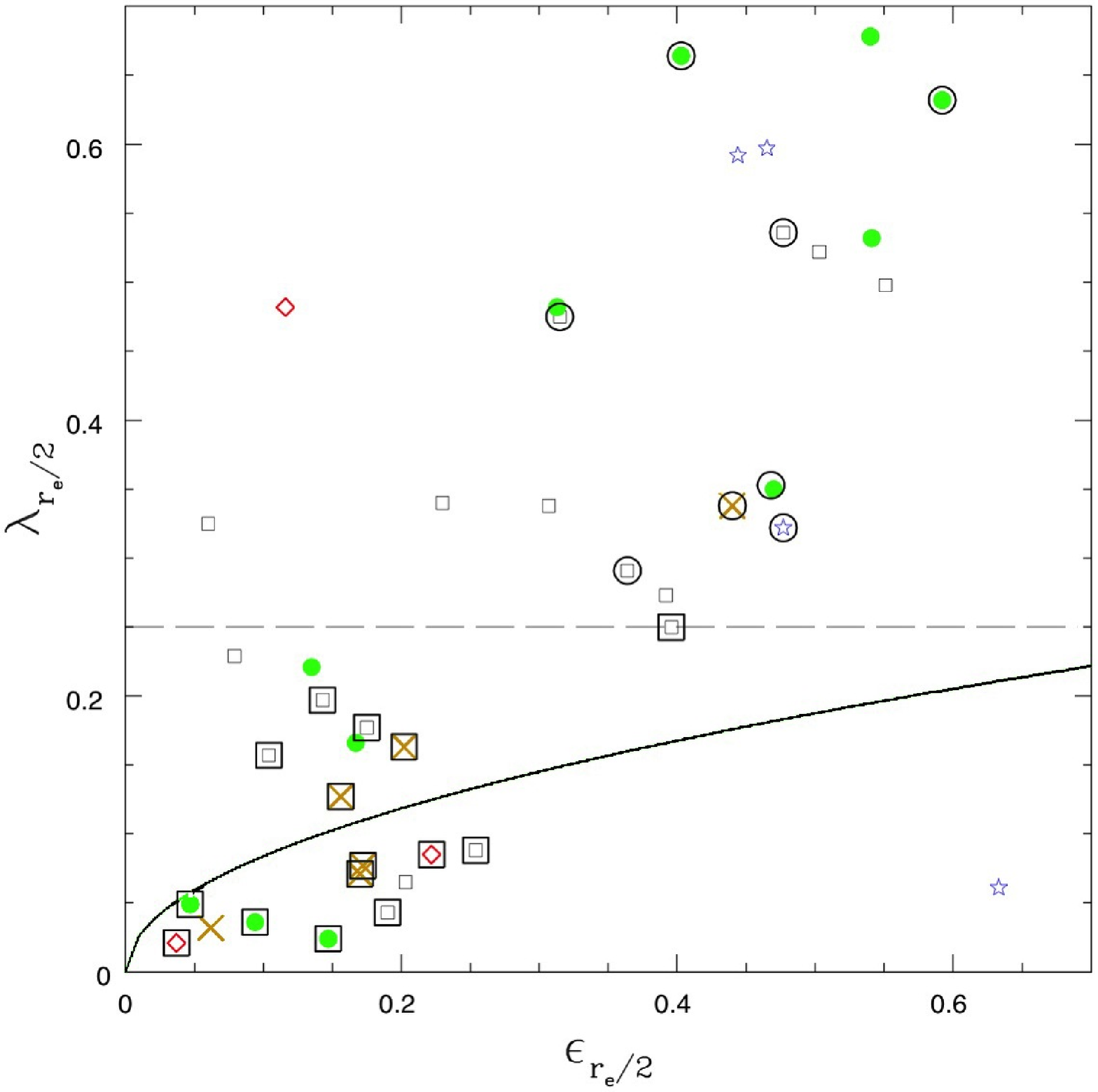}
\caption{  Specific baryonic angular momentum, $\lambda_{r_e/2}$,   
vs. average ellipticity, $\epsilon_{r_e/2}$, measured for the 41 ETGs in Table~2
(see references in the table caption.)  The solid line separates FR from SR according to \citet{Emsellem11}.
The dashed line separates ETGs with {\it cuspy} profiles (large circles) from  
{\it core} galaxies (large squares) in the sample of \citet{Lauer12}.
MIR classes are coded as follows: square = class~0; 
cross = class~1; full dot = class~2; star = class-3 and diamonds = class~4. } 
\label{fig3}
\end{figure*}
%-------------------------- end figure3  ----------------------------------------------------

\subsection{The ACS Virgo and Fornax  ETGs samples}

The passive evolution of all the 6 fainter, {\it cuspy} ETGs in our subset of the 
\citet{Kormendy09} sample prompted us to enlarge the sample to fainter galaxies.
A large HST study of the inner regions of ETGs in the Virgo and Fornax clusters
has been performed by \citet{Cote06} and \citet{Turner12}, the ACSVCS and ACSFCS respectively. 
The final sample with MIR spectra includes 23 ETGs in Virgo (top part of Table~2) and 
8 ETGs in Fornax (bottom of Table~2) that have been classified by \citet{Turner12}  
with an ACS class of either, {\it core} (originally non-nucleated)
 or {\it cuspy} (originally nucleated). 
Additional ETGs in low density environments, 
with a MIR spectrum in \citet{Rampazzo13} and a Sersic decomposition
of the luminosity profile, come from \citet{Krajnovic13b} and are reported in
Table~2.  

In Figure~\ref{fig2}  we plot the absolute K$_s$-band magnitude 
distribution of {\it cuspy} and {\it core} ETGs in the Virgo and Fornax clusters
(bottom left panel) and of the entire sample (bottom right panel).
\citet{Cote06} and \citet{Turner12} noticed that virtually none of the bright 
ETGs are {\it cuspy}, in contrast to the fainter ones, as shown by the lower panel of
Figure~\ref{fig2}. The percentage of {\it cuspy} ETGs in Virgo and Fornax  
reaches 67\%$\pm$8\% and 72\%$\pm$13\% respectively among fainter
galaxies in their sample.
In the top left panel of Figure~\ref{fig2} we show the distribution of MIR classes
in the same bins of M$_{K}$ magnitude. Most of the {\it cuspy} ETGs have MIR 
class-0 spectra, indicating passive evolution in their nucleus. Considering
ETGs with M$_{K}\geq$-24, 82$^{+18}_{-16}$\%\footnote{ The fractions reported consider 
upper and lower   errors corresponding to 1$\sigma$ Gaussian errors \citep{Gehrels86}.}  
of the galaxies are {\it cuspy}, of
which 59$^{+25}_{-18}$\% are passively evolving and only 18$^{+17}_{-10}$\%
show PAH features.  
The brightest (M$_{K}\leq$-25) ETGs have {\it cores},
and various MIR spectra are found, from class-0 to  class-4, as 
noticed before in the \citet{Kormendy09} sample.
However, active MIR classes, from 1 to 4, are distributed along all the
 M$_{K}$ range examined, suggesting that ``wet'' accretions occur at all
 magnitude/mass bins.
 
In the right panels of Figure~\ref{fig2} we consider the entire sample of 
44 ETGs with a MIR spectrum and a {\it core}/{\it cuspy} classification. 
Of the whole sample, 62$^{+22}_{-15}$\%  of the passive ETGs 
are {\it cuspy}. For the subset fainter than M$_{K}=-24$, 87$^{+23}_{-18}$\% are 
{\it cuspy}, 57$^{+22}_{-16}$\% are passively evolving and
33$^{+18}_{-12}$\% show PAHs in their MIR spectra.
Although  the increase in number of galaxies is modest
going from cluster to low density environments,  the 
number of active galaxies (classes 1, 2, 3 and 4) tends to 
increase at all magnitude/mass bins, suggesting a possible environmental
effect \citep[see e.g.][]{Clemens06,Clemens09}. E.g. \citet{Serra12} detected
HI in 40\% of ETGs outside Virgo and 10\% inside it. 
   
\section{MIR spectra vs. fast and slow rotators}

 With the aim of investigating the MIR properties of FRs vs. SRs in the  
   $r_e/2$ region, we cross-correlate  the samples adding a measure of 
the specific angular momentum to our sample with a MIR nuclear
 classification. 41 out of 91 ETGs in the \citet[][Table 5]{Rampazzo13} 
 sample of MIR spectra  have a measure of the specific angular momentum. 
 Table~2 provides the list of  ETGs with a FR/SR classification within $r_e/2$ 
(column 8 from \citet[][ATLAS$^{3D}$: Table B1]{Emsellem11}) and their  
MIR class (column 10). The list includes 29 FRs and 12 SRs.
The Fornax ETGs lack  the kinematical  classification in the  
   $r_e/2$ region. 

In Figure~\ref{fig3} the specific angular momentum at $r_e/2$, $\lambda_{r_e/2}$,
is plotted against the corresponding average ellipticity, $\epsilon_{r_e/2}$,
(Table~\ref{tab2}, columns 5 and 6). According to \citet{Lauer12} the dashed line
in the figure separates the {\it core} and {\it cuspy} ETGs into different rotation classes,
leaving the {\it core} set contaminated only by face-on {\it cuspy} galaxies. The
figure shows the 23 galaxies of \citet{Lauer12}, among the 
41 ETGs in \citet{Emsellem11} (large squares for SR, large
circles for FR). Figure~\ref{fig3} plainly shows that different MIR classes are 
found associated with both FR and SR, as well as in {\it core} and {\it cuspy}  ETGs. 

25$^{+24}_{-14}$\% of SRs  and 52$^{+17}_{-13}$\% of FRs have a class-0 spectrum
i.e. are passively evolving. On the other hand, 
of those MIR classes showing emission lines (1, 2, 3 and 4), i.e.
the complement of passive ETGs,   75$^{+25}_{-24}$\% of SRs and  
48$^{+17}_{-12}$\% of FRs are gas rich in their nucleus. 
In the whole sample of \citet{Rampazzo13}, 64$^{+12}_{-6}$\% 
of the nuclei  show emission lines, although with different intensity.
This results are in agreement with optical studies, in which, 
depending on the sample the ionized gas is detected in  $\approx$50-90\% of ETGs \citep{ph86,mac96,Sarzi06,Sarzi10,yan06,serra08,Annibali10}. 
This suggests  that {\it wet-mergers/accretions} play
a role in the recent few Gyrs history of both SR and FR. 

50$^{+30}_{-20}$\% of SRs and 38$^{+15}_{-11}$\% of FRs show 
PAH features (MIR classes 2, 3, 4), indicating that in about half of both SR and FR
a star formation episode has recently occurred \citep{Kaneda08,Panuzzo11,Rampazzo13}.
Galaxies with normal PAH ratios, class-3 spectra, are a minority:  
only 10$^{+8}_{-5}$\% of FR show star forming spectra and 8$^{+19}_{-2}$\%  SR.
All MIR class-4 spectra, 17$^{+22}_{-11}$\% of SR and  3$^{+8}_{-1}$\% of FRs,  show 
PAHs. The optical study of 65 ETGs of \citet{Annibali10} provided 
indications that the AGN phenomenon is associated with star formation. 

\section{Summary and Conclusions}

This note investigates whether the photometric segregation between 
{\it cuspy} and {\it core} nuclei and/or the dynamical segregation between 
fast and slow rotators can be attributed to formation via ``wet'' or ``dry'' mergers.

We explore the question by comparing the MIR spectral characteristics
with their {\it cuspy}/{\it core} morphology 
\citep[][]{Kormendy09,Lauer12,Cote06,Turner12,Krajnovic13a}
and  FR/SR characterization (ATLAS$^{3D}$). We use {\it Spitzer}-IRS spectra
and MIR classes discussed in \citet{Panuzzo11} and \citet{Rampazzo13}.
{\it These spectra are sensitive to the recent few Gyr ($z\lesssim0.2$) 
evolution of the ETGs.}
We find the following:

\begin{itemize}

\item{With the exception of NGC~4365, which is passively evolving, MIR spectra of all the 
bright  {\it core}  ETGs in the  \citet{Kormendy09} sample show nebular
emission lines, and PAH features are detected in 5 out of 9 objects. 
These types of nuclei should have recently accreted gas-rich material.
 If such objects formed via ``dry'' mergers, the process was completed before
 $z\sim0.2$ and ``wet'' accretions have happened since.
%incompatibly with the hypothesis of a dry merger scenario, suggested for the formation of
%their {\it core} nuclei. 
 AGN feedback does not prevent a late star formation episode in the bright Es NGC~4374, NGC~4636 
and NGC~4552. The few (6) faint, {\it cuspy} ETGs in the \citet{Kormendy09} sample
all show passively evolving spectra, irrespective of their magnitude.}

\item{MIR spectra of the total {\it cuspy/core}  sample (44 ETGs),  
confirm that ETGs fainter than M$_{K_s}$=$-24$ mag, mostly {\it cuspy},
are predominantly passively evolving.  This fact is particularly significant in Virgo and Fornax 
where 82$^{+18}_{-10}$\% are {\it cuspy}, {\it the majority also FRs}. 
\citet{Kormendy09} noticed that {\it cuspy} ETGs have disky (positive 
$a_4/a$ Fourier coefficient) isophotes in the nuclear region, a structure 
that suggests some form of dissipation during the formation. 
The passive MIR spectra suggest that either this infall was sterile, i.e.
without star formation,  or happened at $z\gtrsim0.2$, so as to leave no 
trace in the present MIR spectra.
\citet{Khochfar11} models suggest that a fraction of FR lying in high--density environments  
have a residual gas fraction below 15\%, i.e. they have exhausted their cold gas reservoir 
and have no hot halo from which gas can cool. Counterparts in low
density environments (Figure~\ref{fig2} right panel) show the tendency 
to be more gas rich and hence more active.} 

\item{A significant fraction of  both FR (38$^{+18}_{-11}$\%) and 
SR (50$^{+34}_{-21}$\%) shows PAH features in their MIR spectra. 
Ionized and molecular gas are also commonly detected. 
Recent star formation episodes are not a rare phenomenon in either  FR or SR,
even in those dominated by AGN activity \citep[see also.][]{Annibali10}.
Recently, observing HI rich ETGs, \citet{Serra14} found that SRs are detected as 
often, host as much H I  and have a similar rate of H I discs/rings as FRs.}  

\end{itemize}

Despite the expectation that the signature of ``wet'' or ``dry'' merger is strongest in the 
galaxy nucleus, the nuclear MIR spectra do not clearly link the {\it core} versus {\it cuspy} 
morphology and the FR versus SR kinematical class to these alternative formation
scenarios. Within the last few Gyrs, only at the two extremes of the ETG luminosity, 
does the dichotomy emerge: the brightest {\it core}, mostly SR  and faint  {\it cuspy}, mostly FR
(in the Virgo and Fornax  clusters) separate into mostly active and passive ETGs,
 respectively. The result, however, is  in contrast to what is expected for {\it core}-SR
versus  {\it cuspy}-FR, i.e. a  ``dry'' versus ``wet'' accretion scenarios. 

The obvious possibility is that these photometric and  kinematical classes are signatures 
generated by  the two different evolutionary scenarios at $z\gtrsim0.2$  
\citep[see e.g.][]{Khochfar11},  so that they do not  affect the MIR spectra. 
On the other hand,  adopting the traditional E/S0s morphological subdivision 
\citet[][]{Rampazzo13} found that Es are significantly more passive than S0s in the
same epoch. FR/SR classes may smooth away differences between 
Es and S0s, since a large fraction of Es transit into the FR class. 
At the same time, recent observations tend to emphasize the complexity of 
ETGs when their study is extended to large radii \citep[see also][]{Serra14}. 
\citet{Arnold14} recently obtained the extended,  up to 2-4 $r_e$,
 kinematics of  22 ETGs in RSA. They find that only SRs remain slowly rotating 
 in their halos, while the  specific angular momentum of ETGs classified as FR 
 within $r_e=1$, may dramatically change at larger radii.  
 \citet{Arnold14} suggest that the traditional E/S0  classification better 
 accounts  for the observed kinematics up to large radii and likely of their
 complex evolutionary scenario.
 
\section*{Acknowledgements}

RR acknowledges partial financial support by  contracts ASI-INAF I/016/07/0 and 
ASI-INAF I/009/10/0.  OV acknowledges support from the Conacyt grant CB-2012-183013.
This research has made use of the NASA/ IPAC Infrared Science Archive, which is operated by the Jet Propulsion Laboratory, California Institute of Technology, under contract with the National 
Aeronautics and Space Administration.

%\appendix

%\section[]{}

\end{document}